\documentclass[aps,prl,twocolumn,groupedaddress,amsfonts,showpacs]{revtex4}
\usepackage{amsmath,amssymb,bm,graphicx,epsfig,psfrag}

\usepackage[normalem]{ulem} %%required for various underlining styles

\newcommand{\Prm}{\mathrm{Pr}_\mathrm{m}}   %magnetic Prandtl
\newcommand{\Rey}{\mbox{\rm Re}}            %Reynolds
\newcommand{\Rm}{R_\mathrm{m}}              %magnetic Reynolds
\newcommand{\Rmc}{R_\mathrm{m,cr}}          %critical magnetic Reynolds
\newcommand\sfrac[2]{{\textstyle{\frac{#1}{#2}}}}

\begin{document}

\title{Magnetic structures produced by the small-scale dynamo}

\author{S.~Louise~Wilkin}
\author{Carlo~F.~Barenghi}
\author{Anvar~Shukurov}
\affiliation{School of Mathematics and Statistics, Newcastle University,
Newcastle upon Tyne, NE1 7RU, UK}

\date{\today}

\begin{abstract}
Small-scale dynamo action has been obtained for a flow
previously used to model fluid turbulence, where the sensitivity of the
magnetic field parameters to the kinetic energy spectrum can be explored. We
apply quantitative morphology diagnostics, based on the Minkowski functionals,
to magnetic fields produced by the kinematic small-scale dynamo to show that
magnetic structures are predominantly filamentary rather than sheet-like.  Our
results suggest that the thickness, width and length of the structures scale
differently with magnetic Reynolds number as $\Rm^{-2/(1-s)}$ and
$\Rm^{-0.55}$ for the former two, whereas the latter is independent of $\Rm$,
with $s$ the slope of the energy spectrum.
\end{abstract}


\maketitle

The fluctuation (or small-scale) dynamo is a turbulent dynamo mechanism where a random
flow of electrically conducting fluid generates a random magnetic field with zero mean
value (the other type of dynamo relies on deviations of the flow from
mirror symmetry, and perhaps other symmetries,
and produces mean magnetic fields).
This dynamo mechanism appears to be responsible for the random magnetic
fields in the interstellar medium \citep{BBMSS96} and in galaxy clusters
\citep{Setal05,SSH06,EV06}.
 A necessary condition for the small-scale dynamo is
that the magnetic Reynolds number $\Rm$ is large enough that the random velocity shear
dominates over the effects of the fluid's electric resistivity. Here $\Rm=ul/\eta$
where $u$ and $l$ are a typical velocity and length-scale respectively and $\eta$ is
the magnetic diffusivity. The critical
magnetic Reynolds number, $\Rmc$,
is $30$--$500$, depending on the nature of the flow
\cite{Kazantsev,ZRS83,RogachevskiiIK97,BoldyrevSC04,SchekochihinHBCMMc05,BS05}.
Numerical simulations of the nonlinear, saturated states of the small-scale dynamo have
been reviewed in ref.~\cite{BS05}, but more insight can be gained by a
careful analysis of the kinematic regime where the magnetic field is still too weak to affect
%the flow which generates it. In particular, the turbulent inertial range is undeveloped
the flow. In particular, the turbulent inertial range is undeveloped
for the modest values of the Reynolds number achievable in direct numerical
%simulations; therefore, numerical results are often ambiguous or difficult to interpret.
simulations; therefore, numerical results are often ambiguous.

Here we study the kinematic small-scale dynamo by solving numerically the induction equation
\begin{equation}\label{indeq}
\frac{\partial{\bf{B}}}{{\partial{t}}}= \nabla
\times \left( {\bf{u}} \times {\bf{B}} \right) +\eta \nabla^2 {\bf{B}}, \qquad
\nabla \cdot {\bf B} = 0,
\end{equation}
for the magnetic field $\mathbf{B}$ in a prescribed flow $\mathbf{u}(\mathbf{x},t)$.
Unlike other numerical models of the small-scale dynamo, our choice of the velocity
field allows us to control fully its energy spectrum and time variation.
In particular, the flow has a well pronounced power-law
spectral range with
controllable spectral slope.
We use a
spectral model for $\mathbf{u}$, specified in Eq.~(\ref{uF}), which was developed as
a Lagrangian model of turbulence and contains its essential features: it is three
%dimensional, time-dependent, multi-scale, exhibits any desired energy spectrum and
dimensional, time-dependent, multi-scale and
has transport properties (two-particle dispersion, etc.) which agree remarkably well
with those of turbulent flows \cite{Malik99,Fung98} in both experiments and
numerical calculations.
Therefore, our main results are comparable with numerical experiments \cite{HBD04,SCTMM04}
and to some extent with experimental results \cite{VKS07}.
The velocity field ${\bf u}$ is given by
\begin{equation}\label{uF}
{\bf u}({\bf x},t)= \sum_{n=1}^{N}\left[{\bf C}_n \times \widehat{\bf k}_n
\cos \phi_n + {\bf D}_n \times\widehat{\bf k}_n \sin \phi_n \right],
\end{equation}
where
$\phi_n={\bf k}_n \cdot {\bf{x}} + \omega_n t$,
$N$ is the number of modes and $\widehat{\bf k}_n$ are randomly chosen unit
vectors (${\bf k}_n= k_n \widehat{\bf k}_n$ is the wave vector); note that $\bf{u}$ is
solenoidal by construction. The \emph{directions\/} of ${\bf C}_n$ and ${\bf D}_n$ are
chosen randomly; we require, however, that they are normal to ${\bf k}_n$, so that
the root mean velocity of each mode
is $\lbrack(C_n^2+D_n^2)/2\rbrack^{1/2}$.
The magnitudes of $\mathbf{C}_n$ and $\mathbf{D}_n$ are chosen to reproduce
the desired energy spectrum, $E(k)$:
$C_n= D_n =[\frac23 E(k_n) \Delta k_n]^{1/2}$, where
$\Delta k_n=\frac12(k_{n+1}- k_{n-1})$ for
$2\leq n \leq N-1$, but $\Delta k_1=\frac12(k_2-k_1)$ and $\Delta k_N=\frac12(k_N-k_{N-1})$.
For example, a model spectrum with
suitable behaviors at $k\to0$, $k\to\infty$,
and $E(k)\propto k^s$ within the inertial range
$k_0\ll k\ll k_\mathrm{d}$, can be implemented with
\begin{equation}\label{Ekn}
   {E(k_n)} = a k_n^4 \left[ 1+
%        \left(\frac{k_n}{k_{0}} \right)^2 \right]^{\frac{s-4}{2}}
        \left(\frac{k_n}{k_{0}} \right)^2 \right]^{(s-4)/2}
        \exp\left[-{\textstyle\frac{1}{2}}
                \left(\frac{k_n}{k_{\mathrm{d}}}\right)^2\right],
\end{equation}
where $k_0$
is the integral scale of the flow, $k_\mathrm{d}$ is the
dissipation wavenumber and $a$ is the normalization constant used to control the intensity of
the flow. The Kolmogorov inertial range spectrum is obtained for $s=-5/3$.
The frequencies $\omega_n=[k_n^3 E(k_n)]^{1/2}$ introduce time variation such
that each mode varies at its `eddy turnover' time.

Having specified $k_0$ and $k_\mathrm{d}$, we can introduce the effective
Reynolds number via
$\Rey=(k_\mathrm{d}/k_0)^{(1-s)/2}$
and
$\Rm= \langle u^2 \rangle ^{1/2}l_0/\eta$ for the magnetic Reynolds number,
where angular brackets denote suitable averaging and $l_0=2 \pi /k_0$.
%
% FIGURE 1-------------------------------------------------------
\begin{figure}
      \includegraphics[width=0.33\textwidth]{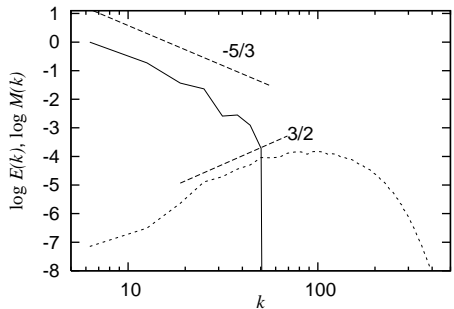} %{tryUandB.eps}
      \caption{\label{fig1} Typical energy spectra for the velocity field ($E$,
solid) and a growing magnetic field ($M$, dotted) in arbitrary units, with
$N=40$, $k_0=2 \pi$ and $k_{\mathrm{d}} \approx 16 \pi$; this corresponds to
the effective Reynolds number $\Rey \approx 16$ and $\Rm \approx 2000$. Dashed
lines represent power laws with the slopes indicated.}
\end{figure}
%-----------------------------------------------------------------
%
We enforce periodicity in the unit box by choosing ${\bf k}_n$ randomly from a
family of vectors whose components are multiples of $2 \pi$.
The resulting kinetic energy spectrum $E(k)$ for $s=-5/3$ is shown in
Fig.~\ref{fig1} (solid line), defined as
$\int_0^\infty E(k) \, dk =V^{-1} \int_V \sfrac{1}{2}|{\bf u}|^2 \, dV,$
where $V$ is total volume, as obtained numerically via Fourier transform of $\bf{u}$
given by Eq.~\eqref{uF} with the input from Eq.~\eqref{Ekn}.
Although the flow is not random, it exhibits Lagrangian chaos \cite{Malik99,Fung98},
which facilitates the dynamo action. The flow is stationary on average, so we expect that
the mean magnetic energy density will vary exponentially in time, $\langle
B^2\rangle^{1/2}\propto e^{\sigma t}$,
with the growth (or decay) rate $\sigma$ depending on $\Rm$ and the kinetic energy spectrum
\cite{RogachevskiiIK97}. Meanwhile, the spatial form of the magnetic field is expected to
remain statistically stationary.

% FIGURE 2--------------------------------------------------------------
\begin{figure}
   \begin{center}
      \psfrag{sigmagrowthratess}{growth rate, $\sigma$}
      \psfrag{Rm}{$\Rm$}
      \includegraphics[width=0.33\textwidth]{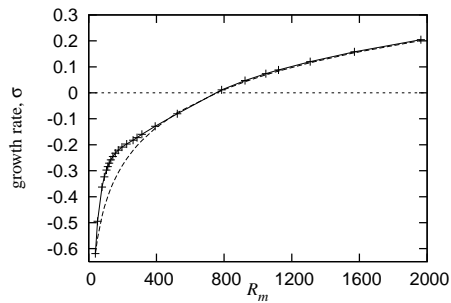} %{si_fitpa.eps} %{sigma_v_Rm.eps}
      \caption{\label{fig2} The growth (or decay) rate of the r.m.s.\ magnetic field
versus the magnetic Reynolds number for a
flow shown in Fig.~\ref{fig1} with $\Rey\approx16$;
note that $\sigma=0$ for $\Rm={\Rmc}\approx 753$. The dashed curve
is an asymptotic fit as discussed in the text.}
   \end{center}
\end{figure}
%----------------------------------------------------------------------

We solve the induction equation (\ref{indeq}) in a unit cubic periodic box
discretized in $128^3$ grid cubes. We begin with a periodic, seed magnetic field with
energy distributed equally at $5$ different scales in the box. After a small number of
timesteps, the $5$-peaked magnetic energy spectrum smooths out to become a horizontal flat
spectrum. Figure~\ref{fig2} shows $\sigma$, measured in terms
of the turnover time at the smallest velocity scale, $2\pi/\omega_n$.
Dynamo action is initiated at
$\Rmc \approx 753$ for the Kolmogorov spectrum, $s=-5/3$. The data were obtained by changing
$\Rm$ (or $\eta$) without changing the flow. Thus, the effective magnetic Prandtl number
$P_\mathrm{m}=\Rm/\Rey$ varies together with $\Rm$.  Also plotted (dashed curve) is the
asymptotic behavior suggested for $\Rm \gg 1$ in ref.~\cite{RogachevskiiIK97}:
$\sigma \approx \alpha (\cal{U}/\cal{L})\,\ln\left(\Rm/\Rmc\right)$,
where $\cal{U}$ and $\cal{L}$ are a representative speed and length respectively and
$\alpha$ is a constant of order unity; \cite{RogachevskiiIK97} obtain $\alpha \approx
2/3$ for ${\cal{U}}=u_0$, ${\cal L}=2 \pi/k_0$. A very good fit to our results is
achieved for a similar value $\alpha \approx 0.6$ (or $\approx
0.2$ if the dissipation scale is used). The accuracy of this fit for $\Rm >
\Rmc$ is over $96 \%$. Magnetic structures at the largest values of
$\Rm$ and $k$ considered here may be only marginally resolved with the
resolution $128^3$ (cf.\ ref.~\cite{Mininni06}) but note that
the size of our computational box is unity rather than $2\pi$. However, the systematic
behavior of the growth rate and magnetic spectrum at the largest values of
$\Rm$ and $k$ indicate that the numerical resolution is sufficient; we
carefully checked for other signs of insufficient resolution in our results
but none were found.

The energy spectrum $M(k)$
of the growing magnetic field (defined similarly to $E(k)$) for
$\Rm \approx 2000$ is shown in Fig.~\ref{fig1} (dotted line), where it exhibits a
range with $M(k)\propto k^{3/2}$ expected for a small-scale dynamo with a
single-scale random flow \cite{Kazantsev}. As we discuss below, our
model also reproduces all other known features of the small-scale dynamo that we
have tested.

\begin{figure}
   \begin{center}
      \includegraphics[width=0.33\textwidth]{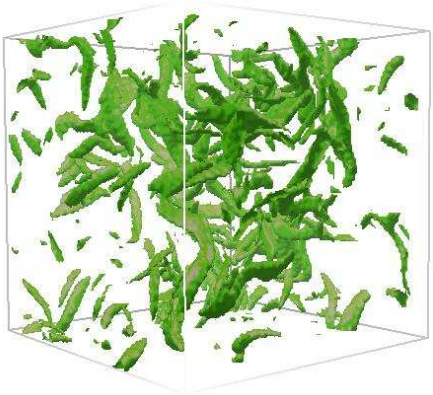} %{dynamo_iso.eps}
      \caption{\label{fig3} (color online)
The $B^2=3\langle B^2\rangle$ isosurfaces of magnetic energy density
for the flow of Fig.~\ref{fig1} with  $\Rm=1570$.
}
   \end{center}
\end{figure}

Asymptotic solutions of the induction equation for $\Rm \gg 1$ relevant to the small-scale dynamo
have been interpreted as indicating that magnetic field is mostly concentrated into
filaments of thickness $l_\eta=l_0 \Rm^{-1/2}$ (for single-scale flows)
\cite{ZRS83} or $l_\eta=l_0 \Rm^{-2/(1-s)}$ for flows with kinetic energy spectral
index
$s$ \cite{RogachevskiiIK97,Vincenzi02,BoldyrevSC04}.
However, more recent numerical simulations appeared to display
magnetic sheets and ribbons \cite{SchekochihinHBCMMc05,BS05,Mininni06}.
The latter conclusion was based on the visual
inspection of magnetic isosurfaces and application of heuristic morphology
indicators. Here we apply mathematically justified morphology quantifiers, based on
the Minkowski functionals \cite{ArmstrongMA83}, to isosurfaces of magnetic energy density
such as those shown in Fig.~\ref{fig3}.
This tool has previously been applied to galaxy distribution and cosmological structure formation
\cite{Sahni98,Schmalzing99,Schmalzing97}.

% FIGURE 3---------------------------------------------------------------
%-------------------------------------------------------------------------

\begin{figure}[t]
    \begin{center}
      \includegraphics[width=0.44\textwidth]{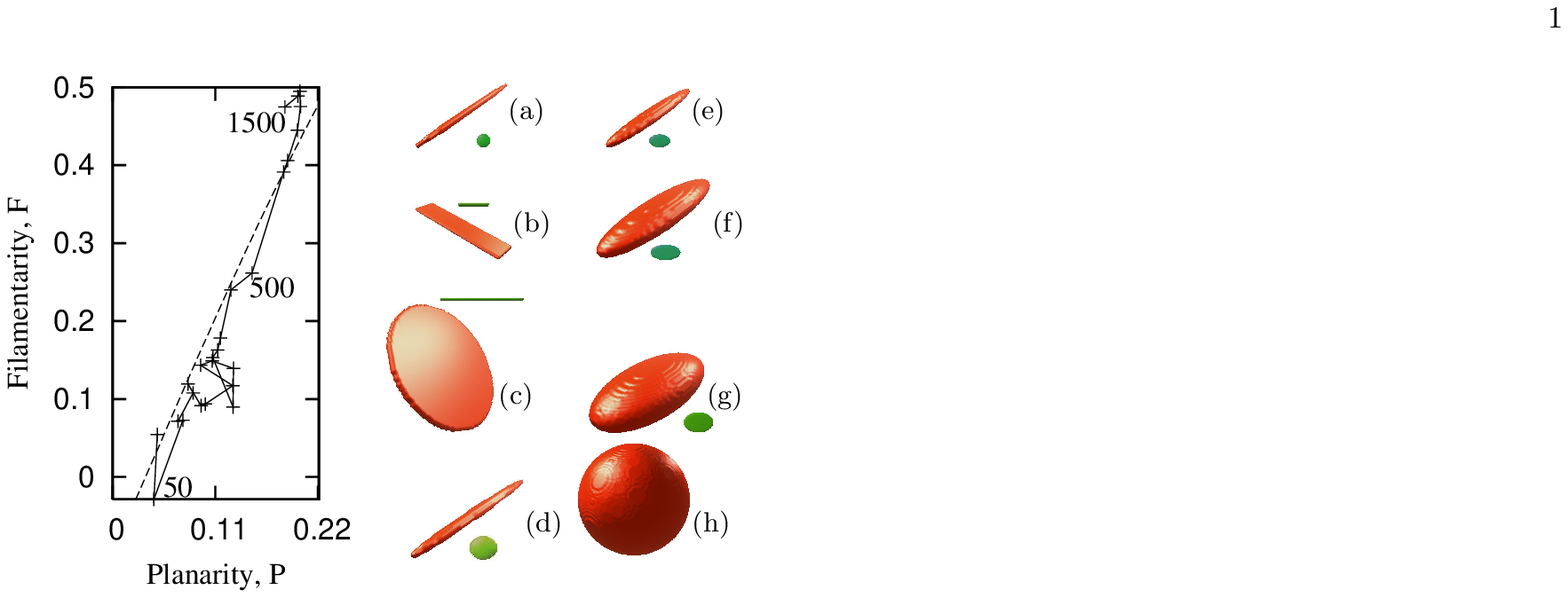}
      \caption{\label{fig4} (color online) The route taken by magnetic structures
through the
$(P,F)$-plane as $\Rm$ increases, averaged for each $\Rm$ between eight isosurfaces
ranging from $B^2=1.5\langle B^2\rangle$ to $B^2=5\langle B^2\rangle$ in
steps of $0.5\langle B^2\rangle$; in this range the
magnetic structures depend only weakly on the choice of isosurface.
Three data points are labelled with their values of $\Rm$. The dashed line
is the estimates of Eq.~\eqref{PFestimate}. The objects on the right,
from a filament to a sphere, are shown for reference:
{\bf{$(P,F)=$}}
(a) $(0.096,0.81)$;
(b) $(0.66,0.23)$;
(c) $(0.66,0.12)$;
(d) $(0.25,0.66)$;
(e) $(0.18,0.43)$;
(f) $(0.14,0.23)$;
(g) $(0.087,0.073)$ and
(h) $(0.0036,-0.0047)$.
The cross section (in green) is also displayed
for each object, except for (h) where it is circular.}
  \end{center}
\end{figure}

The morphology of structures in three dimensions can be fully quantified using the four
Minkowski functionals:
\begin{equation}
\begin{tabular}{l@{}l}
$V_0=\displaystyle\iiint  dV$\;,&  \hspace{2.5mm}$V_1=\sfrac{1}{6}\displaystyle\iint dS$\;, \\
\\[2pt]
$V_2=\displaystyle\frac{1}{6 \pi}\iint(\kappa_1 +\kappa_2 )\, dS$\;,&
                                   \hspace{2.5mm}$V_3=\displaystyle\frac{1}{4 \pi} \iint\kappa_1 \kappa_2 \,dS$\;, \\
\end{tabular}
\end{equation}
where integration is over the volume and surface of the structures,
respectively, and $\kappa_1$ and $\kappa_2$ are the principal curvatures of the
surface.  The Minkowski functionals can be used to
calculate the typical thickness, width and length of the structures, as $T = V_0/2V_1$,
$W = 2V_1/\pi V_2$, and $L = 3V_2/4V_3$, respectively. Then, useful dimensionless measures of
`planarity' $P$, and `filamentarity' $F$ \cite{Sahni98} can be defined as
$P=(W-T)/(W+T)$, $F= (L-W)/(L+W)$.
In idealized cases and for convex surfaces, values of $P$ and $F$ lie between zero
and unity. For example, an infinitely thin pancake has $(P,F)=(1,0)$, a perfect
filament has $(P,F)=(0,1)$, whereas $(P,F)=(0,0)$ for a sphere.  A $B^2=1.5
\langle B^2 \rangle$ isosurface of the
initial magnetic field has $(P,F)=(0.094,0.14)$. Some other examples are
shown on the right-hand side of Fig.~\ref{fig4}.
We note that
the unit cube has $T=3/4$, $W=2/\pi$, $L=1/2$, thus these measures are not necessarily
such that $T<W<L$. Deviations from this ordering are relatively
rare for random fields studied here, yet to avoid confusion we introduce
the notation $l_1={\rm min}(T,W,L)$, $l_2={\rm med}(T,W,L)$ and $l_3={\rm max}(T,W,L)$.

% FIGURE 4------------------------------------------------------------------
%----------------------------------------------------------------------

As shown in Fig.~\ref{fig4}, $F$ increases faster than $P$ with $\Rm$, so the
magnetic structures produced by the small-scale dynamo are better described as
filaments, especially at the larger values of $\Rm$.
Remarkably, similar velocity field structures are not filamentary since
the wave vectors in Eq.~(\ref{uF}) have no preferred direction, hence
the velocity field at small scales is nearly isotropic.
Correspondingly, the isosurfaces of ${\bf u}^2$ have negligible planarity
and filamentarity. The isosurface of vorticity $\mathbf{\Omega}$ with
$\Omega^2=4\langle\Omega^2\rangle$ has $(P,F)=(0.18,0.11)$;
similarly the isosurface of the total strain ($S^2=S_{ij}S_{ij}$) with
$S^2=4.5\langle S^2 \rangle$ has $(P,F)=(0.11,0.16)$. Thus, the
morphology of the magnetic field is controlled by the nature of the dynamo
action rather than by immediate features of the velocity field.
Isosurfaces of the electric current density
$\mathbf{J}=\nabla\times\mathbf{B}$ are ribbon-like, with $(P,F)=(0.57,0.82)$
at a level $J^2=4\langle J^2\rangle^2$ for $\Rm=1570$.

\begin{figure}
   \includegraphics[width=0.33\textwidth]{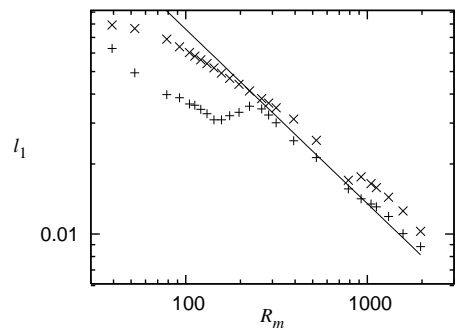}
\caption{\label{fig5} Average values of the thickness $l_1$
of the isosurface structures against $\Rm$ at two instants in time for the flow
of Fig.~\ref{fig1}. Data marked `$\times$' (`+')
are results attained when the smallest eddies have made 25 (50) revolutions.
The solid straight line has a gradient of $-3/4$.}
\end{figure}

Using the Minkowski functionals, we can also reliably measure the characteristic
length scales of magnetic structures and explore their scalings with $\Rm$ and $s$.
In Fig.~\ref{fig5}, we display $l_1$ against $\Rm$ at two instants in time for the
flow with $s=-5/3$ shown in Fig~\ref{fig1}.
Whilst the behavior for $\Rm \ll \Rmc$ shows variations in time, we
observe for $\Rm \gtrsim 200$ a time-independent scaling of the thickness of
magnetic structures: $l_1 = 2 \pi/k_\eta \sim \Rm^{-3/4}$.
This scaling, obtained in ref.~\cite{RogachevskiiIK97} for a flow
$\delta$-correlated in time (see also
ref.~\cite{Vincenzi02,BoldyrevSC04}) follows from the balance of the induction and
diffusion terms in the induction equation \cite{Subramanian99}: a flow with
energy spectrum $E(k)\propto k^s$ yields
\begin{equation}\label{eq:scale}
l_1\simeq l_0 \Rm^{-2/(1-s)}.
\end{equation}
%
%
% FIGURE 5---------------------------------------------------------------
%------------------------------------------------------------------------------
The reason for the $\Rm$-dependence of this %characteristic
scale is akin to that of the magnetic diffusion length, or the skin-depth.

\begin{figure}
   \includegraphics[width=0.41\textwidth]{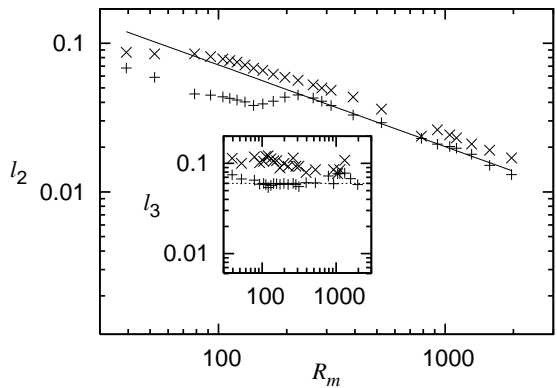}  %{lWLvRmmultiplot.eps}
\caption{\label{fig6} As in Fig.~\ref{fig5}, but for the width $l_2$ (main
frame) and length $l_3$
(inset). The solid straight line in the main plot has a gradient of $-0.55$.}
\end{figure}

Figure ~\ref{fig6} shows the
characteristic width $l_2$ and length $l_3$ of the magnetic structures
against $\Rm$. For $\Rm \gtrsim 200$ we observe another time-independent scaling, $l_2
\sim \Rm^{-0.55}$. This distinct behavior of the width of magnetic structures
has not been obtained in earlier analytical or numerical studies of the
small-scale dynamo. The simultaneous decrease of $l_2$ and $l_1$, coupled with the
approximately $\Rm$-independent behavior of $l_3$ (Fig.~\ref{fig6} inset)
supports the notion that the magnetic structures become filamentary as
$\Rm$ increases (see Fig.~\ref{fig4}).
Indeed, using $l_1 \sim 2.4 \Rm^{-0.75}$,
$l_2 \sim 0.9 \Rm^{-0.55}$ and $l_3 \sim 0.05 \Rm^0$ in
the definitions of $P$ and $F$, we can estimate, for $s=-5/3$,
\begin{equation}\label{PFestimate}
P \sim 1 - 2[1 + {\textstyle\frac{3}{8}} \Rm^{0.2}]^{-1}, \quad F \sim 1-
2[1+ {\textstyle\frac{1}{18}} \Rm^{0.55}]^{-1},
\end{equation}
so that $F > P$ for $\Rm\gtrsim  200$.
These relations are shown by the dashed line in Fig.~\ref{fig4}.

To investigate how the scaling laws identified via the Minkowski functionals compare with
those inferred from other measures, we calculated the inverse `integral scale' of
the magnetic field, $2\pi/l_I = \int{k M(k)}\,dk /\int{M(k)}\,dk$. We found that $l_I$
follows a
scaling of $\Rm^{-0.42}$, which understandably differs from the scalings of $l_1$,
$l_2$ and $l_3$. The scale $l_I$ is a poor measure of the dimension of anisotropic
magnetic structures such as filaments. We note that the above scaling of $l_I$ is
maintained for \emph{all} subcritical and supercritical values of $\Rm$, unlike the results of
Figs.~\ref{fig5} and \ref{fig6} which display well-defined, time-independent
scalings only for $\Rm \gtrsim 200$. We have verified that the
scaling~\eqref{eq:scale} emerges for $s=-5/3$, $-2$, $-3$. On the contrary, $l_2
\sim \Rm^{-0.55}$ independently of $s$. Asymptotic solutions \cite{RogachevskiiIK97}
suggest that the small-scale dynamo (with $\Prm < 1$) is only possible for
$s<-3/2$. Our results show that,
for high effective $\Prm$, the dynamo action is possible for $s=-1$ as
well, although a scaling different from \eqref{eq:scale} is exhibited.
In this context, we propose that the nature of the asymptotic solution, rather than
the possibility of a dynamo, is different at $s=-1$ from that at $s<-3/2$.
We also tested the proxy dimensions of ref.~\cite{Schekochihin04} but these do
not show physically justifiable scalings.
%
% FIGURE6-----------------------------------------------------------------------
%--------------------------------------------------------------------------------

The scaling of $l_1$ with $\Rm$ has been obtained
\cite{RogachevskiiIK97,Vincenzi02,BoldyrevSC04} for flows with
$\Prm=\Rm/\Rey \lesssim 1$, where the maximum of the magnetic spectrum occurs in the
inertial range of turbulence. With the effective $\Rey$ defined above,
our simulations have an effective $\Prm >1$. We cannot introduce a true $\Rey$ because we do not solve
the Navier-Stokes equation and therefore the kinematic viscosity is not defined in our
model. The reason why scalings expected for $\Prm \lesssim 1$ occur in our model which
appears to have $\Prm >1$ remains unclear.

In conclusion, we have shown that velocity fields of the form (\ref{uF}) are
able to initiate the small-scale dynamo. This model flow reproduces many
properties of turbulence and thus offers a convenient tool to study
dynamos in multi-scale, well controlled flows. A
quantitative, mathematically justifiable description of the morphology of
magnetic structures based on the Minkowski functionals indicates that, at
least at the kinematic stage, magnetic field is concentrated into filaments
rather than sheets or ribbons. We have confirmed that the characteristic thickness of
magnetic structures scales with the magnetic Reynolds number as
given in Eq.~(\ref{eq:scale}) for the energy spectrum slope of $s$, whereas the
width varies as $\Rm^{-0.55}$ independently of $s$, and the length is independent of $\Rm$.

\begin{acknowledgments}
We acknowledge useful discussions with S.~Tobias and K.~Subramanian. This work was
supported by the EPSRC grant PB/KH/037026987 and the Leverhulme Trust grant F/00
125/N.
\end{acknowledgments}

%\bibliography{bibfile}

\end{document}